\documentclass{aa}
\usepackage{graphicx}
\usepackage{txfonts}
\begin{document}
\title{The physical structure of the point-symmetric and \\
       quadrupolar planetary nebula NGC 6309}

\author{R. V\'azquez\inst{1} \and
              L. F. Miranda\inst{2} \and
              L. Olgu\'{\i}n\inst{1,3} \and
              S. Ayala\inst{3,4} \and \\
              J. M. Torrelles\inst{5} \and
              M. E. Contreras\inst{1} \and
              P. F. Guill\'en\inst{1}}

\offprints{R. V\'azquez}

\institute{
Instituto de Astronom\'{\i}a, Universidad Nacional Aut\'onoma de
M\'exico, Apdo. Postal 877, 22800 Ensenada, B. C., Mexico\\
\email{vazquez, mcontreras, fguillen@astrosen.unam.mx} \and
Instituto de Astrof\'{\i}sica de Andaluc\'{\i}a, CSIC, P. O.
Box 3004, E-18080 Granada, Spain \\ \email{lfm@iaa.es} \and
Instituto de Astronom\'{\i}a, Universidad Nacional Aut\'onoma de
M\'exico, Apdo. Postal 70-264, 04510 M\'exico, D. F., Mexico\\
\email{lorenzo@astrosen.unam.mx} \and
Centro de Radioastronom\'{\i}a y Astrof\'{\i}sica, Universidad
Nacional Aut\'onoma de M\'exico, Apdo. Postal 3-72 (Xangari),
Morelia, Mich., Mexico\\ \email{s.ayala@astrosmo.unam.mx}  \and
Instituto de Ciencias del Espacio (CSIC) - IEEC, Facultat de
F\'{\i}sica, Universitat de Barcelona, Av. Diagonal 647, 08028
Barcelona, Spain\\ \email{torrelles@ieec.fcr.es}}

\date{Received ---  / Accepted --- }

\abstract{}
{We analyse the point-symmetric planetary nebula \object{NGC\,6309}
in terms of its three-dimensional structure and of internal
variations of the physical conditions to deduce the physical
processes involved in its formation.} {We used VLA-D
$\lambda3.6$-cm continuum, ground-based, and HST-archive imaging
as well as long slit high- and low-dispersion spectroscopy.}
{The low-dispersion spectra indicate a high excitation nebula,
with low to medium variations of its internal physical conditions 
 ($10,600\,{\rm K}     \lesssim T_{\rm e}$[\ion{O}{iii}] $\lesssim 10,900\,{\rm K}$; 
  $10,100\,{\rm K}     \lesssim T_{\rm e}$[\ion{N}{ii}]  $\lesssim 11,800\,{\rm K}$; 
  $1440\,{\rm cm}^{-3} \lesssim N_{\rm e}$[\ion{S}{ii}]  $\lesssim 4000\,{\rm cm}^{-3}$;
  $1700\,{\rm cm}^{-3} \lesssim N_{\rm e}$[\ion{Cl}{iii}]$\lesssim 2600\,{\rm cm}^{-3}$;
  $1000\,{\rm cm}^{-3} \lesssim N_{\rm e}$[\ion{Ar}{iv}] $\lesssim 1700\,{\rm cm}^{-3}$).
The radio continuum emission indicates a mean electron density of
$\simeq1900$\,cm$^{-3}$, emission measure of
$5.1\times10^5$\,pc\,cm$^{-6}$, and an ionised mass
$M$(\ion{H}{ii})$\simeq0.07\,$M$_\odot$. In the optical images,
the point-symmetric knots show a lack of [\ion{N}{ii}] emission
as compared with similar features previously known in other PNe.
A rich internal structure of the central region is seen in the HST
images, resembling a deformed torus. Long slit high-dispersion
spectra reveal a complex kinematics in the central region, with
internal expansion velocities ranging from $\simeq20$ to
30\,km\,s$^{-1}$. In addition, the spectral line profiles from the
external regions of NGC\,6309 indicate expanding lobes
($\simeq40$\,km\,s$^{-1}$) as those generally found in bipolar
nebulae. Finally, we have found evidence for the presence
of a faint halo, possibly related to the envelope of the AGB-star
progenitor.} {Our data indicate that NGC\,6309 is a quadrupolar
nebula with two pairs of bipolar lobes whose axes are oriented
PA=${40\degr}$ and PA=${76\degr}$. Equatorial and polar velocities
for these two pairs of lobes are 29 and 86\,km\,s$^{-1}$ for the
bipolar system at PA=${40\degr}$ and 25 and 75\,km\,$s^{-1}$ for
the bipolar system at PA=${76\degr}$. There is also a central torus
that is expanding at 25\,km\,s$^{-1}$. Kinematical age for all
these structures is around 3700 to 4000\,yr. We conclude that
NGC\,6309 was formed by a set of well-collimated bipolar outflows
(jets), which were ejected in the initial stages of its formation
as a planetary nebula. These jets carved the bipolar lobes in the
previous AGB wind and their remnants are now observed as the
point-symmetric knots tracing the edges of the lobes.}

\keywords{planetary nebulae: individual: NGC 6309
               --  ISM: kinematics -- ISM: abundances }

\titlerunning{The physical structure of NGC\,6309}
\authorrunning{V\'azquez et al.}
\maketitle

\begin{figure*}
\centering
  \includegraphics[width=7.0in]{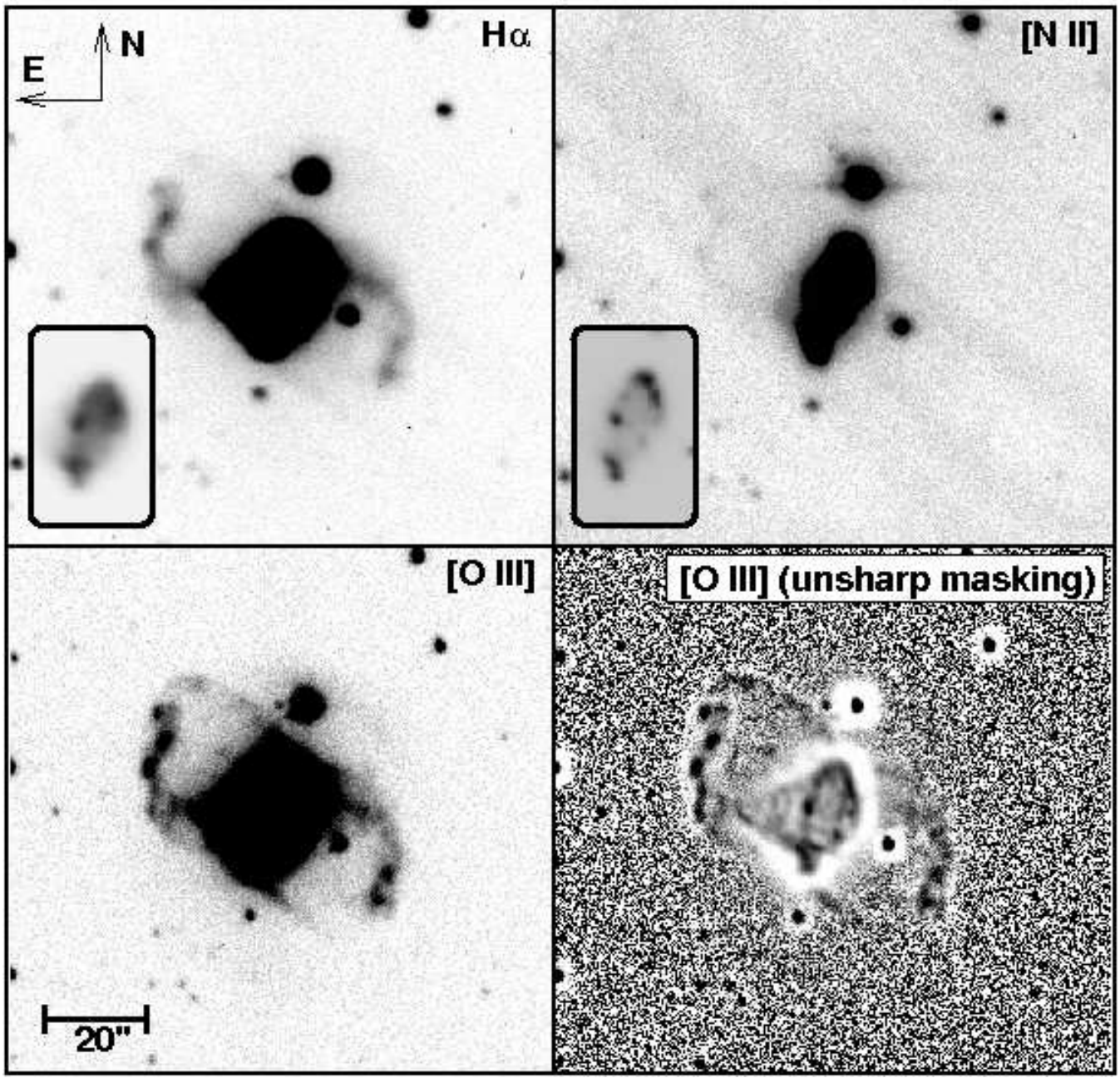}
  \caption{Images of NGC\,6309 in the light of H$\alpha$ (top left),
[\ion{N}{ii}] (top right) and [\ion{O}{iii}] (bottom left), including
an unsharp masking of the [\ion{O}{iii}] image (bottom right). North
is up and east is left in all the panels. Small frames are included
inside the upper panels to show the structure of the central region.}
  \label{cafos}
\end{figure*}

\section{Introduction}

In spite of all the work that up to now has been done to understand
the origin of planetary nebulae (PNe) morphologies and their
evolution (e.g., Kwok, Purton \& FitzGerald \cite{kwo78};
Kahn \& West \cite{kah85}; Balick \cite{bal87}; Icke \cite{ick88};
Mellema \cite{mel95}; Perinotto et al. \cite{per04}; Rijkhorst,
Mellema \& Icke \cite{rij05}; Sch\"onberner et al. \cite{sch05};
Sch\"onberner et al. \cite{sch07}), some morphological structures
still remain as unsolved problems. In particular, the so-called
point-symmetry (Stanghellini, Corradi \& Schwarz, \cite{sta93};
Gon\c calves et al. \cite{gon03}) appears enigmatic.
Some observational studies about point-symmetric PNe (e.g.
Miranda \& Solf \cite{mir92}; L\'opez, Meaburn \& Palmer
\cite{lop93}) relate the formation of these objects to collimated
outflows from a precessing central source known as bipolar rotating
episodic jets (BRETs, see L\'opez, V\'azquez \& Rodr\'{\i}guez
\cite{lop95}). However, in some cases, there are not definite proofs
about the jet nature of these features as they are not confirmed
by spectroscopic studies (e.g. V\'azquez et al. \cite{vaz99a},
V\'azquez et al.\cite{vaz02}). Although some theoretical models
have intended to explain point-symmetric PNe (e.g., Cliffe et al.
\cite{cli95}; Livio \& Pringle \cite{liv96}, \cite{liv97};
Garc\'{\i}a-Segura \& L\'opez \cite{gar00}; Rijkhorst, Icke
\& Mellema \cite{rij04}), the origin and shaping of this kind of
objects remains a puzzle.

NGC 6309 is a PN whose morphology strongly suggests a BRET
scenario for its origin, as can be seen in the
H$\alpha$+[\ion{N}{ii}] and [\ion{O}{iii}] images by Schwarz,
Corradi \& Melnick (\cite{sch92}). It has two prominent,
point-symmetric `arms' formed by pairs of condensations in
addition to a bright internal elliptical structure. Some previous
studies on \object{NGC\,6309} have determined an expansion
velocity $V_{\rm exp}$[\ion{O}{iii}]=34\,km\,s$^{-1}$
(Sabbadin \cite{sab84}), as well as mean physical parameters
and total abundances (G\'orny et al. \cite{gor04}), namely,
electron density and temperature 
$N_{\rm e}$[\ion{S}{ii}]=2600\,cm$^{-3}$,
$T_{\rm e}$[\ion{N}{ii}]=12,097\,K, $T_{\rm
e}$[\ion{O}{iii}]=11,845\,K; and
abundance ratios He/H=0.1, N/H=8.20, O/H=8.64, Ne/H=7.82,
and S/H=6.49 (log H=12, except for He/H ratio).
Armour \& Kingsburgh (\cite{arm01}) found similar values.
NGC\,6309 has also been included in the list of PNe with
low ionisation structures (LIS) by Gon\c calves et al.
(\cite{gon03}). Finally, the central star of \object{NGC\,6309}
has been classified as a ``weak emission line star'' by
G\'orny et al. (\cite{gor04}). 

In spite of this suggestive morphology, an internal kinematic
study of NGC\,6309, as well as a full analysis of the physical
conditions in the different nebular regions, has not yet been done.
In this paper, we have carried out a radio-optical study of NGC\,6309,
including radio continuum mapping, optical imaging, and long-slit
optical spectroscopy, in both, high- and low-dispersion. We discuss
our results on the morphology, kinematics, physical conditions, ionic
abundances, and nature of the gas emission to explore the possible
mechanisms involved in the formation of NGC\,6309.

\section{Observations and results}

\subsection{CCD ground-based imaging}

We obtained narrow-band direct images with the Calar Alto Faint Object
Spectrograph (CAFOS) in the 2.2\,m telescope at Calar Alto Observatory
(CAHA\footnote{The Centro Astron\'omico Hispano-Alem\'an (CAHA) at
Calar Alto, is operated jointly by the Max-Planck Institut f\"ur
Astronomie and the Instituto de Astrof\'{\i}sica de Andaluc\'{\i}a
(CSIC).}) in 1998 July 8. We used two filters centered in H$\alpha$
($\Delta\lambda=15${\AA}) and [\ion{N}{ii}]6583 ($\Delta\lambda=20${\AA})
as well as a CCD Loral with $2048\times2048$ pixels. The exposure time
for both images was 1800\,s. The scale was 0\farcs33\,pixel$^{-1}$ and
the seeing was about 1\farcs7. We obtained an additional image on 2004
August 3 with the 1.5\,m telescope at the Observatorio de Sierra
Nevada \footnote{The Observatorio de Sierra Nevada is operated by
the Consejo Superior de Investigaciones Cient\'{\i}ficas through the
Instituto de Astrof\'{\i}sica de Andaluc\'{\i}a (Granada, Spain)}.
The detector was a RoperScientific VersArray CCD with 2048$\times$2048
pixels, each of 0\farcs232\,pixel$^{-1}$. We used an [\ion{O}{iii}]5007
($\Delta\lambda=50${\AA}) filter with an exposure time of 900\,s.
The images were reduced following standard procedures within the
MIDAS package. In this case, seeing was 1\farcs3.

Figure~\ref{cafos} shows the images of NGC\,6309 in H$\alpha$,
[\ion{N}{ii}] and [\ion{O}{iii}]. An additional unsharp masking
[\ion{O}{iii}] image is also presented in Fig.~\ref{cafos} to show
the link between internal and external nebular structures. In these
images, the high contrast between the brightness of the central region
and that of the `arms' is evident. In particular, the lack of emission
of [\ion{N}{ii}] from the `arms' is unexpectedly remarkable given that,
in general, this kind of point-symmetric microstructures is related to
LIS (Gon\c calves et al. \cite{gon03}). However, there are also some
cases in which the emission of [\ion{N}{ii}] is marginal
(e.g. \object{IC\,5217}; Miranda et al. \cite{mir06}). In the case of
NGC\,6309, the lack of [\ion{N}{ii}] from the point-symmetric knots
indicates that these cannot be considered as LIS.

\begin{figure*}
\centering
  \includegraphics[width=7.0in]{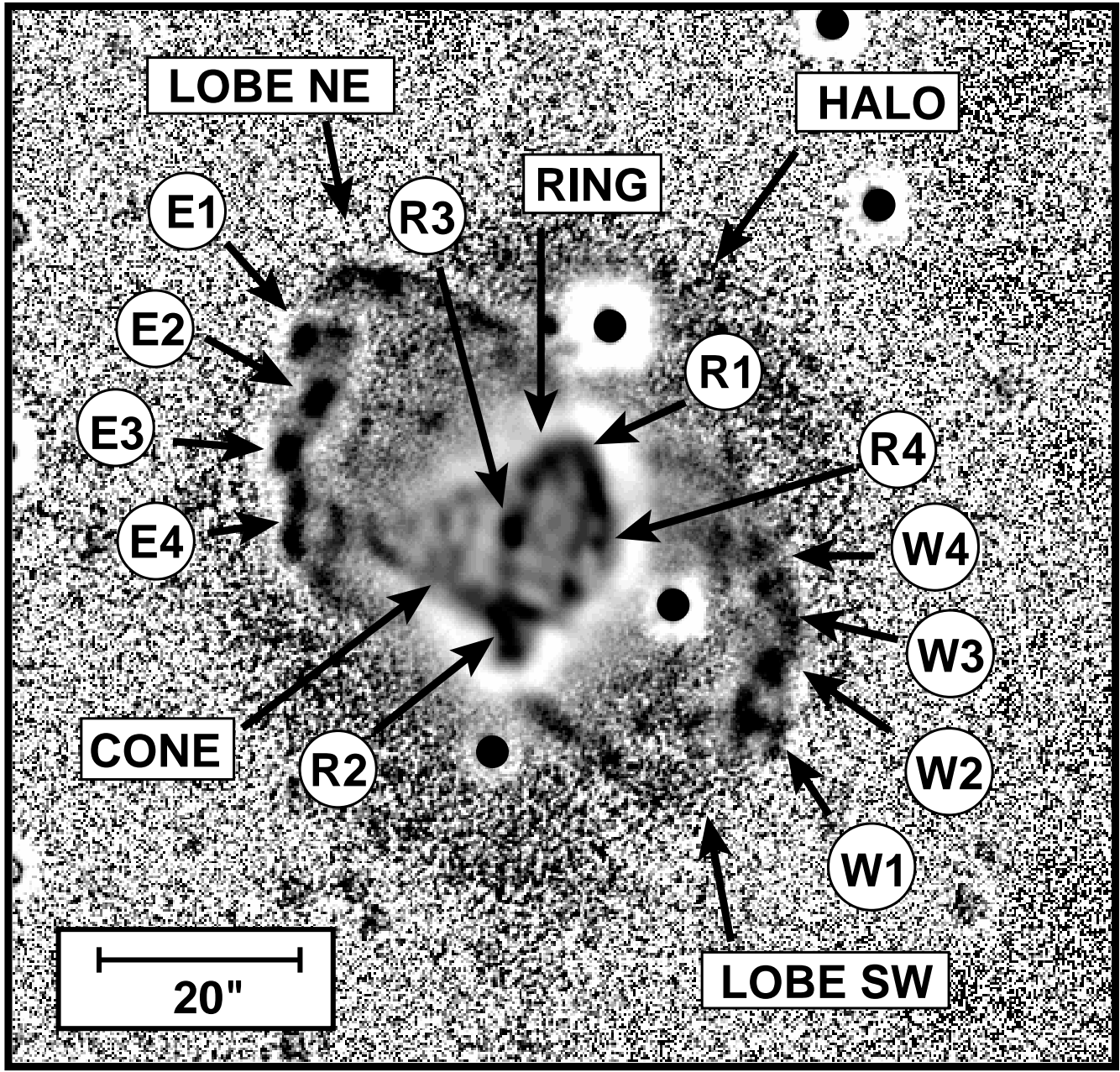}
  \caption{Unsharp-masking image of NGC\,6309 in the light of
[\ion{O}{iii}]. North is up and east is left. Main morphological
features as well as some regions of particular interest are
pointed out and labeled.}
  \label{halo}
\end{figure*}

The structure of the `arms' consists of at least four pairs of
point-symmetric knots, some of them clearly extended perpendicular
to their corresponding radial vector from the central star.
The central region of the NGC\,6309 is a bright ellipse (major axis
$\simeq20\arcsec$, PA\,$-14\degr$) embedded in the main body of the
nebula. The arms appear to leave from the vertexes of this ellipse.
The unsharp masking [\ion{O}{iii}] image also shows a conelike
structure with its base located on the central ellipse and the vertex
on the NE arm. In [\ion{O}{iii}], faint emission connects the arms with
the bright ellipse and seems to trace asymmetrical lobes. In fact, the
appearance of NGC\,6309 in [\ion{O}{iii}] resembles that of a
quadrupolar PN in which one of the outflows (at PA\,40\degr) protrude
into the lobes of the other (at PA\,76\degr). Part of the SW lobe
appears to be open. Finally, a circular faint halo (probably spherical)
is detected in [\ion{O}{iii}] image (Figs.~\ref{cafos} and \ref{halo}).
Its center coincides with the central star of the nebula and its size
is $\simeq56\arcsec$.

In Fig.~\ref{halo}, we show the unsharp masking [\ion{O}{iii}] image
and label some morphological features that will be discussed later.
Knots in the NE arm are named {\bf E1}, {\bf E2}, {\bf E3}, and
{\bf E4}, whereas those in the SW arm are named {\bf W1}, {\bf W2},
{\bf W3}, and {\bf W4}. Regions in the ellipse are called {\bf R1}
and {\bf R2} for the tips of its major axis, and {\bf R3} and
{\bf R4} for those along the minor axis. 

\subsection{HST-WFPC2 imaging}

\begin{figure}
  \resizebox{\hsize}{!}{\includegraphics{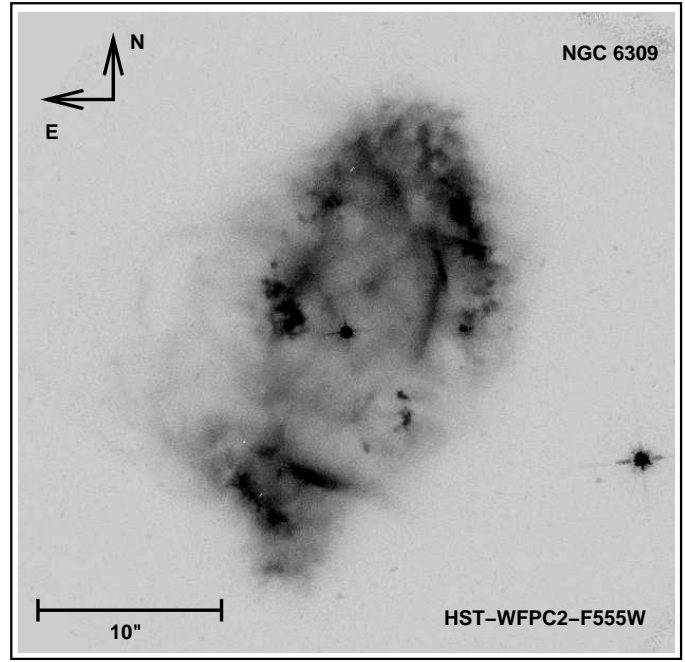}}
  \caption{HST broad-band CCD image of NGC\,6309 in the filter F555W.}
  \label{hst}
\end{figure}

An Hubble Space Telescope (HST) broadband image from the MAST
Archive\footnote{Some of the data presented in this paper were
obtained from the Multimission Archive at the Space Telescope
Science Institute (MAST). STScI is operated by the Association
of Universities for Research in Astronomy, Inc., under NASA contract 
NAS5-26555. Support for MAST for non-HST data is provided by the NASA
Office of Space Science via grant NAG5-7584 and by other grants and
contracts.} was used in order to improve the general view of the
internal morphology of NGC\,6309 (proposal ID: 6119; PI: H. E. Bond;
Date of observation: 1995 August 26). Figure~\ref{hst} shows the
140\,sec image using the filter F555W (nearly Johnson V). This image
can be compared with the optical ground-based images in
Fig.~\ref{cafos}. The bright ellipse is formed by many clumps as well
as diffuse gas. The ellipse is open in its SW and SE regions, whereas
an apparently double structure is observed in the S and N regions. In
addition, a system of faint bubbles is detected towards the NE, which
is probably related to the cone-like structure seen in [\ion{O}{iii}]
(see above). Given the short exposure time of the HST image, evidence
of the ``arms'' is very marginal.

\subsection{$\lambda$3.6-cm VLA-D}

We obtained radio continuum observations at $\lambda$3.6-cm, toward
NGC\,6309 with the Very Large Array (VLA) of the National Radio
Astronomy Observatory (NRAO)\footnote{The National Radio Astronomy
Observatory is a facility of the National Science Foundation operated
under cooperative agreement by Associated Universities, Inc.} in the
D configuration during 1996 August 8. The standard VLA continuum mode
with a 100 MHz bandwidth and two circular polarizations was employed.
The flux and phase calibrators were $1331+305$ (adopted flux density
5.2\,Jy) and $1733-130$ (observed flux density 10.7\,Jy),
respectively. We set phase center at 
$\alpha(2000)=17^{\rm h}14^{\rm m}03\fs6$,
$\delta(2000)=-12{\degr}54{\arcmin}37{\arcsec}$. The on-target
integration time was 28 min. We calibrated the data and processed
it using standard procedures of the Astronomical Image Processing
System (AIPS) package of the NRAO. We obtained a cleaned map of
NGC\,6309 using the task IMAGR of AIPS (parameter ROBUST$=-3$;
Briggs \cite{bri95}). Self-calibration was also performed resulting
in a final synthesized beam of $13\farcs9$ in diameter and rms noise
of the map of $\sigma=33\,\mu{\rm Jy\,beam}^{-1}$.

\begin{figure}
  \resizebox{\hsize}{!}{\includegraphics{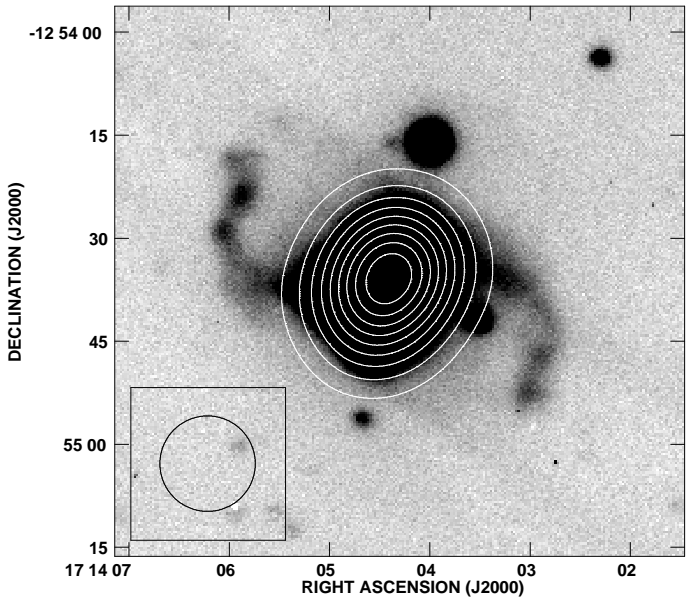}}
  \caption{Contour plot $\lambda3.6$-cm continuum map of NGC\,6309,
overimposed on a H$\alpha$ image. Contour levels are 10, 20, 30, 40,
50, 60, 70, 80, and 90\% of the peak flux ($6.3\times10^{-3}$\,Jy).
The half-power beam width (13\farcs9 in diameter) is shown in the
bottom-left corner.}
  \label{vla}
\end{figure}

The detected radio continuum emission is related to the central
region and appears partially resolved. Based on a 2-D Gaussian fit
(task IMFIT in AIPS), we estimate a size of
{$\simeq19\arcsec\times16\arcsec$} at PA$=-27\degr$ (see
Fig.~\ref{vla}). The position of the intensity peak was found at 
$\alpha(2000)=17^{\rm h}14^{\rm m}04\fs28$, 
$\delta(2000)=-12{\degr}54{\arcmin}37{\arcsec}$. The total flux
density from the radio map is $\simeq115$-mJy, which is similar to
those measured by Milne \& Aller (\cite{mil82}); Ratag \& Pottasch
(\cite{rat91}); and Condon \& Kaplan (\cite{con98}); at 2\,cm
(146-mJy); 6\,cm (102-mJy); and 21\,cm (132-mJy); respectively.
We did not detect emission from the `arms'.

We also derived the mean physical conditions, based on the formalism
of Mezger \& Henderson (\cite{mez67}). Optically-thin emission and 
$T_{\rm e}=10,000$\,K were assumed, obtaining the following results:
$N_e=1900\,{\rm cm}^{-3}$; M(\ion{H}{ii})=$0.07\,M_{\odot}$;
$EM=5.1\times10^5\,{\rm pc\,cm^{-6}}$. We assumed distance to the
nebula to be 2\,kpc as the average value from the different
estimates, which range between 1.1 and 2.5\,kpc (e.g., Daub
\cite{dau82}, Phillips \& Pottasch \cite{phi84}, Amnuel, et al.
\cite{amn84}, Maciel \cite{mac84}, Cahn, Kaler \& Stanghellini,
\cite{cah92}).

\subsection{Long slit low-dispersion spectroscopy}

\begin{figure}
  \resizebox{\hsize}{!}{\includegraphics{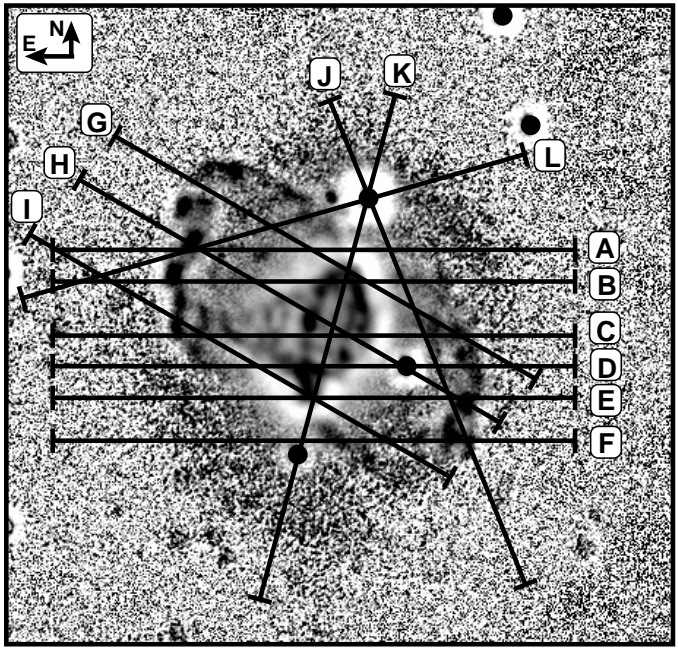}}
  \caption{Slits used in spectroscopy overimposed on [\ion{O}{iii}]
unsharp masking image.}
  \label{rendijas}
\end{figure}

Low-dispersion optical spectra were obtained with the Boller \& Chivens
spectrometer in the 2.1\,m UNAM telescope at the San Pedro M\'artir
Observatory (OAN-UNAM) in 1999 July 19-20 (grating of 300\,lines/mm)
and 2002 August 7-8 (grating of 400\,lines/mm). In all of these cases,
a CCD Tek $1024\times1024$ was used as detector. We set the slit width
to 220\,$\mu$m (1.6\arcsec). The spatial scale is
1\farcs05\,pixel$^{-1}$ whereas the spectral scale is 3 and
4\,{\AA}\,pixel$^{-1}$ for the gratings of 400 and 300\,lines/mm,
respectively. We set the slit at several position angles, which are
labeled as C, J, K, L (grating of 300\,lines/mm), and A, C, and E
(grating of 400\,lines/mm), in Fig.~\ref{rendijas}.

Table~\ref{tboller1} shows the dereddened spectra (considering case
B of recombination) and physical conditions obtained with the 
Five-Level Atom Diagnostics Package ({\it nebular}) from
IRAF\footnote{The Image Reduction and Analysis Facility (IRAF) is
distributed by the National Optical Astronomy Observatories, which
are operated by the Association of Universities for Research in
Astronomy, Inc., under cooperative agreement with the National
Science Foundation.} (applying the extinction law by Howarth
\cite{how83} for dereddening) for the regions {\bf R1}, {\bf R2},
{\bf R3}, {\bf R4}, {\bf E2}, {\bf W1}, and {\bf W1}.

\begin{table*}
\caption{Dereddened spectra and physical conditions from different
regions in NGC\,6309. Intensity values are relative to
$I_{\rm H\beta}$=100. Extinction law ($f_\lambda$) is from Howarth
(1983). The size of the integration region (in arcsec) along the slit
is 7\farcs4 for {\bf R1}, {\bf R2}, {\bf R3}, and {\bf R4}, and
5\farcs3 for the rest. Slit width was set at 1\farcs6 in all the
cases.}
\vskip 0.5cm
{\centering \begin{tabular}{@{}llrrrrrrrrrrr}
            \hline
            \noalign{\smallskip}
Ion               &$\lambda_0$ ({\AA})&$f_\lambda$
                                &  R1 &  R2 &  R3 &  R4 &  E2 &  W1 &  W2 \\
\noalign{\smallskip}\hline\noalign{\smallskip}
{[\ion{O}{ii}]}   & 3727 & 0.256&  108&  102&  ---&  ---&  ---&  ---&  ---\\
{[\ion{Ne}{iii}]} & 3869 & 0.230&  163&  165&  ---&  ---&  ---&  ---&  ---\\
{[\ion{Ne}{iii}]} & 3968 & 0.210&   68&   67&  ---&  ---&  ---&  ---&  ---\\
{[\ion{S}{ii}]}   & 4071 & 0.189&  6.5&  ---&  ---&  ---&  ---&  ---&  ---\\
H$\delta$ + \ion{He}{ii}&4101&0.182&33&   34&  ---&  ---&  ---&  ---&  ---\\
H$\gamma$         & 4340 & 0.127&   55&   53&   26&   27&   52&  ---&  ---\\
{[\ion{O}{iii}]}  & 4363 & 0.121&   12&   12&  7.9&  8.5&  ---&  ---&  ---\\
{\ion{He}{i}}     & 4471 & 0.095&  8.2&  7.1&  3.4&  3.2&  ---&  ---&  ---\\
{\ion{He}{ii}}    & 4540 & 0.078&  ---&  ---&  2.8&  3.3&  ---&  ---&  ---\\
{\ion{N}{iii}}    & 4640 & 0.054&  4.3&  5.0&  5.2&  6.1&  ---&  ---&  ---\\
{\ion{He}{ii}}    & 4686 & 0.043&   29&   31&   81&   79&   86&  106&  ---\\
{[\ion{Ar}{iv}]}  & 4711 & 0.037&  6.0&  5.8&   12&   12&  ---&  ---&  ---\\
{[\ion{Ar}{iv}]}  & 4740 & 0.030&  4.6&  4.6&  9.4&   9.3& ---&  ---&  ---\\
H$\beta$          & 4861 & 0.000&  100&  100&  100&  100&  100&  100&  100\\
{[\ion{O}{iii}]}  & 4959 &-0.024&  487&  464&  335&  365&  341&  364&  451\\
{[\ion{O}{iii}]}  & 5007 &-0.036& 1504& 1466&  993& 1081& 1046& 1097& 1238\\
{\ion{N}{i}}      & 5199 &-0.082&  0.9&  ---&  ---&  ---&  ---&  ---&  ---\\
{\ion{He}{ii}}    & 5411 &-0.133&  2.5&  2.3&  6.2&  6.0&  8.5&  ---&  ---\\
{[\ion{Cl}{iii}]} & 5517 &-0.154&  1.1&  1.0&  0.7&  0.7&  ---&  ---&  ---\\
{[\ion{Cl}{iii}]} & 5537 &-0.157&  1.0&  1.0&  0.7&  0.6&  ---&  ---&  ---\\
{[\ion{N}{ii}]}   & 5755 &-0.195&  1.0&  1.0&  0.6&  ---&  ---&  ---&  ---\\
{\ion{He}{i}}     & 5876 &-0.215&   13&   12&  7.1&  8.1&  ---&  ---&  ---\\
{[\ion{O}{i}]}    & 6300 &-0.282&$6.8^{\mathrm{a}}$&$7.2^{\mathrm{a}}$&2.5&1.2&
---&  ---& ---\\
{[\ion{S}{iii}]}  & 6312 &-0.283&  ---&  ---&  1.8&  1.7&  ---&  ---&  ---\\
{[\ion{O}{i}]}    & 6364 &-0.291&  1.9&  1.8&  0.8&  0.4&  ---&  ---&  ---\\
{[\ion{Ar}{v}]}   & 6435 &-0.302&  ---&  ---&  1.4&  1.4&  ---&  ---&  ---\\
{[\ion{N}{ii}]}   & 6548 &-0.318&   24&   31&  9.3&  4.7&  ---&  ---&  ---\\
H$\alpha$         & 6563 &-0.320&  285&  285&  283&  284&  284&  284&  284\\
{[\ion{N}{ii}]}   & 6583 &-0.323&   60&   60&   27&   12&  8.8&  ---&  ---\\
{\ion{He}{i}}     & 6678 &-0.336&  3.4&  3.6&  2.4&  2.6&  ---&  ---&  ---\\
{[\ion{S}{ii}]}   & 6717 &-0.342&  8.0&  9.1&  2.9&  1.3&  ---&  ---&  ---\\
{[\ion{S}{ii}]}   & 6731 &-0.344&   11&   12&  4.8&  2.1&  ---&  ---&  ---\\
{[\ion{Ar}{v}]}   & 7006 &-0.380&  ---&  ---&  3.1&  2.7&  ---&  ---&  ---\\
{\ion{He}{i}}     & 7065 &-0.387&  3.1&  3.1&  1.9&  2.0&  ---&  ---&  ---\\
{[\ion{Ar}{iii}]} & 7136 &-0.396&   17&   19&   14&   13&   19&   21&   21\\
{[\ion{Ar}{iv}]}+[\ion{Fe}{ii}]&7170&-0.401&1&---&0.9& 0.3&  ---&  ---&  ---\\
{\ion{He}{ii}}    & 7178 &-0.401&  ---&  ---&  0.9&  0.7&  ---&  ---&  ---\\
{[\ion{O}{ii}]}   & 7320 &-0.419&$5.6^{\mathrm{b}}$&$5.7^{\mathrm{b}}$&2.3&1.4&
---&  ---&  ---\\
{[\ion{O}{ii}]}   & 7330 &-0.420&  ---&  ---&  2.0&  1.3&  ---&  ---&  ---\\
\noalign{\smallskip}\hline\noalign{\smallskip}
$c_{\rm H\beta}$       & &      & 0.96& 0.97& 0.70& 0.80& 0.86& 0.85& 0.97\\
$\log\,I_{\rm H\beta}$ & &      &-13.4&-13.5&-13.2&-13.4&-15.2&-15.3& -15.9\\
\noalign{\smallskip}\hline\noalign{\smallskip}
$T_{\rm e}${[\ion{O}{iii}]}&(K)         &&10\,800 &10\,900 &10\,700 &10\,600 &
---&  ---& ---\\  
$T_{\rm e}${[\ion{N}{ii}]}&(K)          &&10\,300 &10\,100 &11\,800 &  --- &  ---   &---& ---\\
$N_{\rm e}${[\ion{S}{ii}]}&(cm$^{-3}$)  &&1800    &1440    &4000    & 3600 &  ---   &---& ---\\
$N_{\rm e}${[\ion{Cl}{iii}]}&(cm$^{-3}$)&&2100    &2350    &2600    & 1700  &  ---  &---& ---\\
$N_{\rm e}${[\ion{Ar}{iv}]}&(cm$^{-3}$) &&$>1000$&$>1500$&$>1700$   &$>1500$&  ---  &---& ---\\
\noalign{\smallskip}\hline\noalign{\smallskip}
$\log $H$\alpha$/[\ion{N}{ii}]     &&&0.533& 0.499& 0.889& 1.244&---&---&---\\
$\log $H$\alpha$/[\ion{S}{ii}]     &&&1.182& 1.142& 1.565& 1.913&---&---&---\\
$\log [\ion{O}{iii}]5007/$H$\alpha$&&&0.723& 0.711& 0.545& 0.580&---&---&---\\
$\log \lambda6717/\lambda6731$     &&&0.748& 0.792& 0.603& 0.623&---&---&---\\
\noalign{\smallskip}\hline\noalign{\smallskip}
\end{tabular} \par}
\begin{list}{}{}
\item[$^{\mathrm{a}}$] This emission line is blended with [\ion{S}{iii}]6312{\AA}.
\item[$^{\mathrm{b}}$] This emission line is blended with [\ion{O}{ii}]7330{\AA}.
\item[] Mean uncertainties are $\Delta c_{\rm H\beta}=\pm0.02$, 
                               $\Delta T_{\rm e}=\pm500$\,K,
                               $\Delta N_{\rm e}=\pm400$\,cm$^{-3}$.
\end{list}
\label{tboller1}
\end{table*}

Physical conditions can only be determined for the regions of the
central ellipse. Electron temperature and densities obtained from
the high-excitation emission lines appears to be quite uniform
($T_{\rm  e}$[\ion{O}{iii}]=10\,750\,K,
$N_{\rm e}$[\ion{Cl}{iii}]=2350\,cm$^{-3}$,
$N_{\rm e}$[\ion{Ar}{iv}]=1430\,cm$^{-3}$),
with variations within the estimated uncertainties. However,
those physical parameters obtained from low-excitation emission
ions appear as separated in two different regimes
($T_{\rm  e}$[\ion{N}{ii}]=10\,200\,K, and 11\,800\,K;
$N_{\rm e}$[\ion{S}{ii}]=1620\,cm$^{-3}$ and 3800\,cm$^{-3}$).

The dereddened absolute H$\beta$ flux values from regions {\bf R1},
{\bf R2}, {\bf R3}, and {\bf R4} are similar (see
Table~\ref{tboller1}). However, the corresponding logarithmic
extinction coefficients $c_{\rm H \beta}$ spread from 0.70 to 0.97
(mean value including all regions is 0.87), probably due to
differences in the internal dust distribution, as has been found
in other PNe (e.g. V\'azquez et al. \cite{vaz99b}, Lee \& Kwok
\cite{lee05}).

In addition, we present the more common ratios for plasma diagnostics
at the bottom of Table~\ref{tboller1}, which show that these regions
are emitting by photoionisation-recombination processes with minimal
or nonexistent shock-cooling contributions. This could be expected
in the surroundings of a hot luminous star, however there are several
cases of PNe in which shock-cooling processes make an important
contribution to the microstructures emission (see Gon\c calves
\cite{gon03} for a compilation). For comparison, previous
spectroscopic studies of NGC\,6309 by G\'orny et al.
(\cite{gor04})
give values of $c({\rm H}\beta)\simeq0.88$,
$T_e$([\ion{O}{iii}])$\simeq$11\,845\,K, 
$T_e$([\ion{N}{ii}])$\simeq$12\,097\,K, and
$N_e$([\ion{S}{ii}])$\simeq2\,600\,{\rm cm}^{-3}$. 

We have also estimated ionic and elemental abundances for some regions
using IRAF and the ICF ({\it Ionisation Correction Factors}) method
(Kingsburgh \& Barlow \cite{kin94}), although we are aware that
recently, Gon\c calves et al. (\cite{gon06}) have noticed that
elemental abundances derived by using this method are overestimated
if they are obtained from narrow, long-slit spectra. The results
are shown in Tables~\ref{tboller2} and \ref{tboller3}.

\begin{table*}
\caption{Ionic abundances of NGC\,6309. In the case of ions with
more than one transition, a flux-weighted average was performed.
Electron temperature and density values of $T_e=$10\,700\,K and
$N_e=2400$\,cm$^{-3}$ were used on the calculations.}
\vskip 0.5cm
{\centering \begin{tabular}{@{}lcrrrrrrrrrr}
            \hline
            \noalign{\smallskip}
Ion   &  & R1  & R2  & R3  & R4  & E2$^{\mathrm{a}}$ & W1$^{\mathrm{a}}$  & W2$^{\mathrm{a}}$\\ 
\noalign{\smallskip}\hline\noalign{\smallskip}
He$^{+2}$ &                  & 0.02 & 0.03 & 0.07 & 0.06 & 0.06 & 0.09 & ---  \\
He$^{+}$  &                  & 0.09 & 0.09 & 0.05 & 0.06 & ---  & ---  & ---  \\
\noalign{\smallskip}\hline\noalign{\smallskip}
O$^{+2}$  &($\times10^{-4}$) & 4.68 & 4.93 & 2.02 & 3.78 & 3.66 & 3.82 & 4.41 \\
O$^{+}$   &($\times10^{-5}$) & 4.70 & 5.93 & 1.28 & 2.16 & ---  & ---  & ---  \\
O$^{0}$   &($\times10^{-5}$) & 1.13 & 1.29 & 0.27 & 0.25 & ---  & ---  & ---  \\
\noalign{\smallskip}\hline\noalign{\smallskip}
Ar$^{+4}$ &($\times10^{-7}$) & ---  & ---  & 4.26 & 5.83 & ---  & ---  & ---  \\
Ar$^{+3}$ &($\times10^{-7}$) & 8.78 & 9.22 & 12.2 & 19.4 & ---  & ---  & ---  \\
Ar$^{+2}$ &($\times10^{-6}$) & 1.48 & 1.68 & 0.90 & 1.18 & 1.71 & 1.91 & 1.92 \\
\noalign{\smallskip}\hline\noalign{\smallskip}
N$^{+}$   &($\times10^{-5}$) & 1.15 & 1.37 & 0.37 & 0.25 & 0.18 & ---  & ---  \\
N$^{0}$   &($\times10^{-6}$) & 2.18 & ---  & ---  & ---  & ---  & ---  & ---  \\
\noalign{\smallskip}\hline\noalign{\smallskip}
S$^{+2}$  &($\times10^{-6}$) & ---  & ---  & 2.14 & 3.99 & ---  & ---  & ---  \\
S$^{+}$   &($\times10^{-7}$) & 5.29 & 5.87 & 2.07 & 1.31 & ---  & ---  & ---  \\
\noalign{\smallskip}\hline\noalign{\smallskip}
Ne$^{+2}$ &($\times10^{-4}$) & 1.59 & 1.75 & ---  & ---  & ---  & ---  & ---  \\
Cl$^{+2}$ &($\times10^{-7}$) & 1.28 & 1.28 & 0.63 & 0.92 & ---  & ---  & ---  \\
\noalign{\smallskip}\hline\noalign{\smallskip}
\end{tabular} \par}
\label{tboller2}
\end{table*}

Regions {\bf R1} and {\bf R2} are similar in physical conditions as
well as in ionic abundances. In both regions, the abundances of the
three species of oxygen as well as He$^{+}$, Ar$^{+2}$, N$^{+}$,
S$^{+}$, and Cl$^{+2}$ are enhanced with respect to the other
regions. The opposite occurs for Ar$^{+3}$ and He$^{+2}$.

With reference to elemental abundances, these are consistent with
the mean values for PNe and Type I PNe, within uncertainties (see
Table~\ref{tboller3}). The exception is N, which show a deficiency,
more in agreement with less-evolved objects, but consistent with
previous estimates (Armour \& Kingsburgh \cite{arm01}). Given the
values of these abundances, NGC\,6309 would be a Type II PN
(intermediate population), according to the classification of
Peimbert (\cite{pei78}). This is reinforced by the low N/O ratio,
which is in agreement with the expected  value for a low-mass
central star (0.6\,M$_{\sun}$, according to Kwok \cite{kwo00}).
If this interpretation of the abundances is right, NGC\,6309 would
represent another case of a non-Type I PN with bipolar outflows
(see V\'azquez et al. \cite{vaz99a}). This strengthens the idea
pointed out by V\'azquez et al. that some low-mass central stars
may also develop bipolar morphologies, in contrast to the fact
that bipolar PNe are usually associated with relatively massive
central stars (Corradi \& Schwarz \cite{cor95}).

\begin{table*}
\caption{Elemental abundances of NGC\,6309. ICFs were obtained
following Kingsburgh \& Barlow (1994). Comparison with mean values
of All-type PNe, Type I PNe and other objects are shown. Except for
He, all the abundances relative to H are logarithmic values with H=+12.}
\vskip 0.5cm
{\centering \begin{tabular}{@{}lrrrrrrrrrrrr}
            \hline
            \noalign{\smallskip}
Ion   & R1  & R2  & R3  & R4  & G04$^{\mathrm{a}}$ & PNe$^{\mathrm{b}}$  & 
TI\,PNe$^{\mathrm{c}}$ & \ion{H}{ii}$^{\mathrm{d}}$  & Sun$^{\mathrm{e}}$ \\        
\noalign{\smallskip}\hline\noalign{\smallskip}
He/H&$0.11\pm0.01$&$0.11\pm0.01$&$0.12\pm0.01$&$0.12\pm0.01$&
     $0.10\pm0.01$&$0.12\pm0.02$&$0.13\pm0.04$&$0.10\pm0.01$&$0.09\pm0.01$\\
O /H&$8.74\pm0.02$&$8.72\pm0.02$&$8.72\pm0.01$&$8.74\pm0.01$&
     $8.64\pm0.02$&$8.68\pm0.15$&$8.65\pm0.15$&$8.70\pm0.04$&$8.66\pm0.05$\\
N /H&$7.93\pm0.02$&$7.92\pm0.02$&$8.04\pm0.01$&$7.89\pm0.02$&
     $8.20\pm0.12$&$8.35\pm0.25$&$8.72\pm0.15$&$7.57\pm0.04$&$7.78\pm0.06$\\
Ne/H&$8.26\pm0.02$&$8.25\pm0.02$&--- &--- &
     $7.82\pm1.00$&$8.09\pm0.15$&$8.09\pm0.15$&$7.90\pm0.10$&$7.84\pm0.06$\\
Ar/H&$6.40\pm0.10$&$6.42\pm0.10$&$6.54\pm0.01$&$6.51\pm0.01$&
     ---          &$6.39\pm0.30$&$6.42\pm0.30$&$6.42\pm0.04$&$6.18\pm0.08$\\
S /H&$6.74\pm0.01$&$6.73\pm0.01$&$6.84\pm0.01$&$6.87\pm0.01$&
     $6.49\pm0.05$&$6.92\pm0.30$&$6.91\pm0.30$&$7.06\pm0.06$&$7.14\pm0.05$\\
\noalign{\smallskip}\hline\noalign{\smallskip}
\end{tabular} \par}
\begin{list}{}{}
\item[$^{\mathrm{a}}$] Taken from G\'orny et al. (\cite{gor04}).
\item[$^{\mathrm{b}}$] Average for PNe (Kingsburgh \& Barlow \cite{kin94}).
\item[$^{\mathrm{c}}$] Average for Type I PNe (Kingsburgh \& Barlow \cite{kin94}).
\item[$^{\mathrm{d}}$] Average for \ion{H}{ii} regions (Shaver et al. \cite{sha83}).
\item[$^{\mathrm{e}}$] The Sun (Grevesse, Asplund, \& Sauval \cite{gre07}).
\end{list}
\label{tboller3}
\end{table*}

\subsection{Long slit high-dispersion spectroscopy}

We obtained high-dispersion optical spectra with the Manchester
Echelle Spectrometer (MES) in the 2.1\,m telescope (f/7.5) at the
San Pedro M\'artir observatory (OAN-UNAM) during the period of 2001
May 22-23. We used a CCD SITe with $1024\times1024$ pixels was used
as detector, and set the slit width to 150-$\mu$m (1.6\arcsec).
A $2\times2$ binning was used, resulting in a spatial scale of
0\farcs6\,pixel$^{-1}$ and a spectral scale of
0.1\,{\AA}\,pixel$^{-1}$. We centered the spectral range at the
H$\alpha$ emission line. The spectra were wavelength calibrated with
a Th-Ar arc lamp to an accuracy of $\pm2$\,km\,s$^{-1}$. The achieved
spectral resolution, as indicated by the FWHM of the comparison lines,
was 12\,km\,s$^{-1}$. Several spectra were obtained with the slit
oriented E-W (slits A to F on Fig.~\ref{rendijas}). 

We obtained on 2004 July 29-30, a second series of spectra, using
the same telescope and instrument. In this case, the spectra were 
centered on the [\ion{O}{iii}]\,$\lambda5007$ emission line
(slits G, H, I, and K on Fig.~\ref{rendijas}), and the spectral
scale was 0.08\,{\AA}\,pixel$^{-1}$. 

\begin{figure*}
\centering
\rotatebox{0}{
  \includegraphics[height=9in]{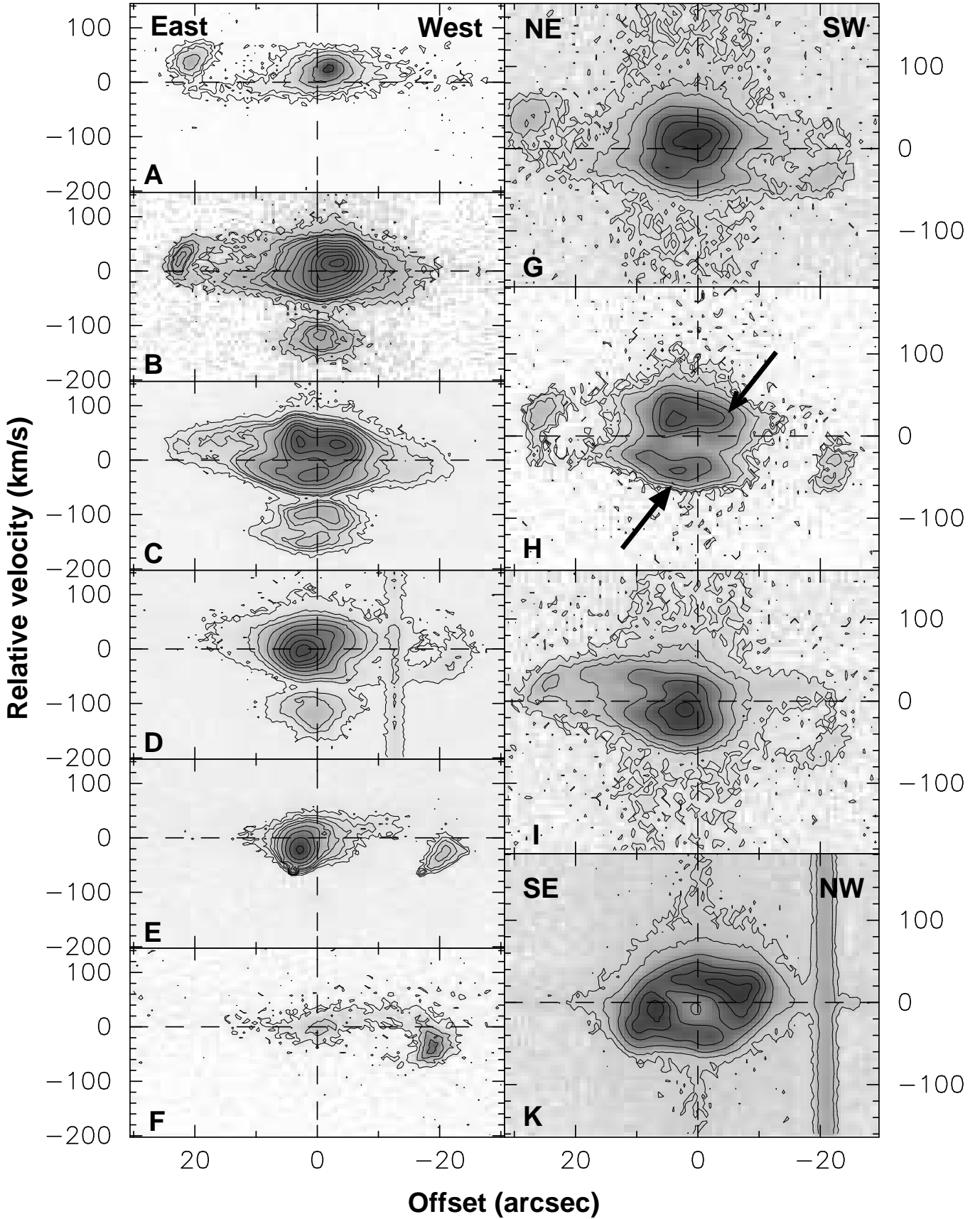}}
  \caption{Position-velocity gray/contour maps corresponding to the
slits A to F in the light of H$\alpha$ (left panels), and slits
G, H, I, and K in [\ion{O}{iii}]\,$\lambda5007$ (right panels).
Slit orientation is indicated, being the same for all the left
panels. In the right panels orientation is indicated on panels
for slits G (valid also for H and I) and K. Velocity scale is
different in left and right panels. The vertical dashed lines
represent the projected position of the central star perpendicular
to the corresponding slit position. On each panel, the horizontal
dashed line corresponds to the systemic radial velocity
($V_{\rm LSR}=-32\pm2$\,km\,s$^{-1}$). Arrows in panel H indicate
location of two features explained in the text.}
  \label{specall}
\end{figure*}

Position-velocity (PV) maps for the H$\alpha$ and
[\ion{O}{iii}]\,$\lambda5007$ are shown in Fig.~\ref{specall}.
We note that the \ion{He}{ii}$\lambda6560$ line is also observed
in the PV maps at the slits B, C, and D. In the PV maps
(Fig.~\ref{specall}) relative position is measured with respect
to the central star whereas radial velocity is measured relative
to the systemic velocity, for which we deduce
$V_{\rm LSR}=-32\pm2$\,km\,s$^{-1}$, based on the expansion velocity
at the position of the central star ([\ion{O}{iii}] spectra from
slits H and K). This value is close to that reported by Schneider
et al. (\cite{sch83}) of
$V_{\rm LSR}\simeq-33.4\pm2.8$\,km\,s$^{-1}$.

In a first step to analyse the high-resolution spectra, we extract
from the PV maps the spatial position and radial velocity of the
individual regions labeled in Fig.~\ref{halo}. The results are listed
in Table~\ref{knots}. The knots in the arms present a noticeable
point-symmetry, both in their position with respect to the central
star and in their radial velocity. In the case of the regions in the
bright ellipse, {\bf R1} and {\bf R2} are moving at the systemic
velocity while {\bf R3} and {\bf R4}, located along the minor axis,
present the maximum radial velocity in this structure. 

In a second step, we analyse the emission-line features in the PV
maps as a whole. Slits H and K are tracing the minor and major axes
of the internal ellipse, respectively. In the PV map of slit H
(Fig.~\ref{specall}), the regions {\bf E2} and {\bf W2} can be
noted as bright condensations located at the extremes of the
emission-line features. Maximum expansion velocity at the lobes
can be measured as $2v_{\rm exp}\simeq80$\,km\,s$^{-1}$ (SW lobe
expansion is best viewed in panel I). The lobe NE shows a bulk
radial velocity of $\simeq+20$\,km\,s$^{-1}$ (as measured at
20{\arcsec} from the center, panel H), whereas the lobe SW
presents $\simeq-20$\,km\,s$^{-1}$ (as measured at 20{\arcsec}
from the center, panel I). The central region also seems to be
expanding, showing a maximum velocity splitting of
$2v_{\rm exp}\simeq69\pm2$\,km\,s$^{-1}$, making it difficult to
determine the radial velocity of the microstructures {\bf R3} and
{\bf R4} accurately. However, the PV map from slit H
([\ion{O}{iii}]) shows evidence of two additional components
separated $\simeq8\arcsec$ and $\simeq46$\,km\,s$^{-1}$ that could
be these features (locations are indicated on Fig. 6 by a pair
of arrows).

Slit K crosses the major axis of the ellipse. In the corresponding
PV map, a velocity ellipse is observed. Features {\bf R1} and {\bf R2}
move with the systemic velocity. However, small regions beyond the
ellipse are also found with radial velocities of
$\simeq \pm 30$\,km\,s$^{-1}$. Maximum expansion velocity is
observed at the center of the ellipse
($2v_{\rm exp}\simeq66\pm2$\,km\,s$^{-1}$), which is nearly compatible
with previous spectroscopic studies (Sabbadin \cite{sab84},
$\simeq68\pm3$\,km\,s$^{-1}$; Armour \& Kingsburgh \cite{arm01},
$\simeq64\pm5$\,km\,s$^{-1}$).

From these slits (H and K), the presence of two expanding asymmetrical
lobes and a velocity ellipse in the center is evident. The expansion
of these bipolar lobes is confirmed with data from the other slits,
and the kinematics of their borders is traced by that of the
point-symmetric knots. 

\begin{table}
\caption{Basic kinematical data for the main morphological features
  in NGC\,6309. Angular distance and position angle (PA) are measured
  from the central star. Radial velocity is measured with respect to
  the systemic velocity (--32\,km\,s$^{-1}$) from the emission line
  indicated in the last column. Uncertainties for the first row
  apply to the rest.}
\vskip 0.5cm

{\centering \begin{tabular}{@{}cr@{}l@{}l@{}l@{}r@{}l@{}l@{}l@{}}
            \hline
            \noalign{\smallskip}
Feature   &  PA     &      &      & Angular     &             & Radial             &$\phantom{X}$& Emission\\
          &  (\degr)&      &      & distance    &             & velocity           &             & line    \\
          &         &      &      & ($\arcsec$) &             & (km\,s$^{-1}$)     &             &         \\
\noalign{\smallskip}\hline\noalign{\smallskip}
 {\bf R1} & --15&$\pm4\degr\phantom{XX}$&9&$\pm3\arcsec$&0    &$\pm2$\,km\,s$^{-1}$&             & [\ion{O}{iii}] \\
 {\bf R2} & +165    &      &  10  &             &   0         &                    &             & [\ion{O}{iii}]\\
 {\bf R3} &  +75    &      &   4  &             & +23         &                    &             & [\ion{O}{iii}]\\
 {\bf R4} &--105    &      &   4  &             &--23         &                    &             & [\ion{O}{iii}]\\
 {\bf E1} &  +50    &      &  30  &             & +35         &                    &             & [\ion{O}{iii}]\\
 {\bf E2} &  +60    &      &  26  &             & +36         &                    &             & H$\alpha$, [\ion{O}{iii}]\\
 {\bf E3} &  +70    &      &  27  &             & +35         &                    &             & H$\alpha$     \\ 
 {\bf E4} &  +90    &      &  25  &             & +24         &                    &             & H$\alpha$     \\
 {\bf W1} &--130    &      &  27  &             &--34         &                    &             & H$\alpha$     \\
 {\bf W2} &--120    &      &  25  &             &--36         &                    &             & H$\alpha$, [\ion{O}{iii}]\\
 {\bf W3} &--110    &      &  24  &             &--34         &                    &             & H$\alpha$, [\ion{O}{iii}]\\
 {\bf W4} & --90    &      &  21  &             &--20         &                    &             & H$\alpha$     \\
\noalign{\smallskip}\hline\noalign{\smallskip}
\end{tabular} \par}
\label{knots}
\end{table}

In the PV map from slit A, we detect an intense emission centered at
2{\arcsec} toward the west and relative velocity centered at
$V_{\rm r}\simeq+20$\,km\,s$^{-1}$. This feature corresponds to the
upper border of the ellipse ($\simeq5\arcsec$ above {\bf R1}). Faint
emission from the halo is detected to the west of this feature (up to
$\simeq10\arcsec - 25\arcsec$) in agreement with the observed size of
this structure. 

On the other hand, the PV map of slit I shows the radial velocity of
feature {\bf E4} (the brightest feature) as $\simeq+24$\,km\,s$^{-1}$,
in addition to the expansion velocity of lobe SW close to
$\simeq40$\,km\,s$^{-1}$. Finally, the PV map of slit C covers the
region at lobe NE corresponding to the cone-like structure. This
region expands to $\simeq25$\,km\,s$^{-1}$, and is located from
5{\arcsec} to 15{\arcsec} toward the East from the center.
The kinematics of the cone is practically indistinguishable from
that of lobe NE.

\section{Discussion}

\subsection{The physical structure of NGC\,6309}

The most reasonable explanation for the morphology and kinematics of
the central ellipse is that such structure corresponds to a tilted
expanding torus. Although other possible structures are possible,
this interpretation appears to be a better fit to our data. Assuming
that the torus is circular and its diameter is $\simeq20${\arcsec}, 
we deduce an inclination angle of the torus axis with respect to the
line of sight of 66{\degr}. The expansion velocity of the torus,
corrected by the inclination angle is
$V_{\rm exp}\simeq25$\,km\,s$^{-1}$. Therefore, the size of this
structure and its kinematic age are $\simeq0.2\,{\rm pc}$ and
$3800\,{\rm yr}$ respectively, assuming a distance of 2 kpc.
In addition, the northeastern half of the torus is blueshifted
whereas the southwestern half is redshifted. We note that these
calculations have been obtained from the earth-based observations.
Since the HST image shows a more complex structure in the torus,
these results should be considered as an approximation only.

\begin{figure}
  \resizebox{\hsize}{!}{\includegraphics{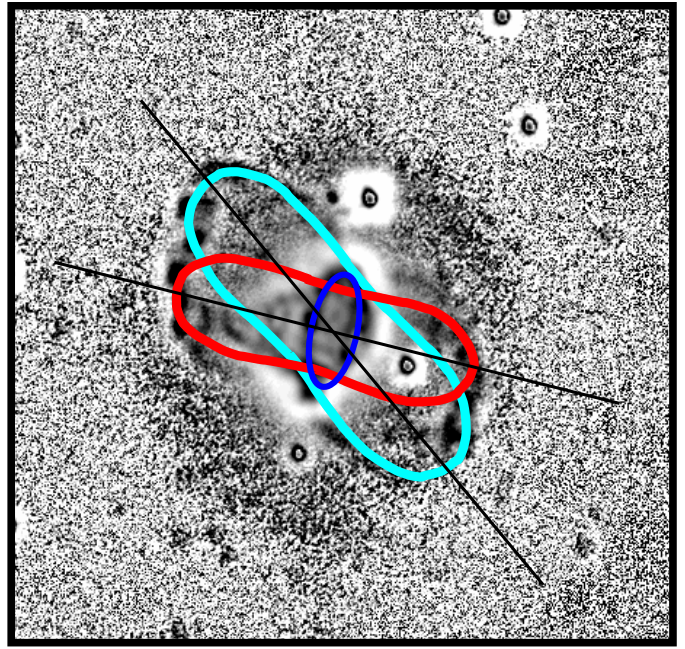}}
  \caption{NGC\,6309 and the two bipolar outflows model. The fit
  of two hour-glass models is overimposed on the [\ion{O}{iii}]
  unsharp-masking image presented before. The central torus is
  also drawn.}
  \label{models}
\end{figure}

With regard to the bipolar lobes, it is clear that a model of 
a single bipolar system with a main axis cannot reproduce the
observed morphology, given the apparent two directions observed
in the nebula. Therefore, we have considered a quadrupolar model
consisting of two pairs of bipolar lobes, one of them oriented at
PA\,$+76\degr$, coincident with the torus axis, and the other at
PA\,$+40\degr$, coinciding with the apparent protrusions in the lobes
along this axis. Each of these bipolar systems is described following
the formulation by Solf \& Ulrich (\cite{sol85}),

\begin{equation}
V(\phi)=V_\mathrm{e}+\left(V_\mathrm{p}-V_\mathrm{e}\right)\times\sin^\alpha(|\phi|),
\end{equation}

\noindent where $\phi$ is the latitude angle above the equator;
$V_\mathrm{e}$ and $V_\mathrm{p}$ are the equatorial and polar
expansion velocity respectively; and $\alpha$ is an exponent that
fits the specific hour-glass shape. With the main assumption that
the axis of the torus and that of the bipolar ejection at
PA\,$+76\degr$ are the same, the kinematics and morphology of this
outflow can be reasonably reproduced with
$V_\mathrm{e}=25$\,km\,s$^{-1}$ (expansion velocity of the torus),
$V_\mathrm{p}=75$\,km\,s$^{-1}$, and $\alpha=6$. There is no clue
about the inclination angle of the outflow at PA\,$+40\degr$. If we
assume the same inclination angle, we find reasonable fits for this
bipolar system with $V_\mathrm{e}=29$\,km\,s$^{-1}$,
$V_\mathrm{p}=86$\,km\,s$^{-1}$, and $\alpha=6$. In Fig.~\ref{models}, 
we show the hour-glass shapes deduced from these fits overimposed on
the unsharp-masking image of NGC\,6309. The observed shape, as well
as the kinematics of the knots, are reasonably reproduced by the
models. With the model data and the assumption of a distance of
2\,kpc, we find a kinematical age of 4\,000\,yr for the bipolar lobes
at PA\,$+40\degr$ and of 3700\,yr for the bipolar system at
PA\,$+76\degr$. These ages are not very different from each other
and are also similar to the age deduced for the torus. This suggests
that the formation of the different structures in NGC\,6309 has
occurred in a relatively small time span. In addition, as these kind
of outflows have been observed in proto-planetary nebulae
(PPN, e.g. Sahai \cite{sah98}), our results impose a minimal value of
time ($\simeq4000$\,yr) for the PPN to PN stage in this nebula. Other
structures such as the ``cone'' and the halo, are present in this
nebula, but we must conduct further studies to understand their
physical properties.  

\subsection{The formation of NGC\,6309}

The appearance of NGC\,6309, as well as the position of the
point-symmetric knots perpendicular to a radial position vector
from the central star, strongly resemble the point-symmetric
structures observed in \object{Fleming\,1} (L\'opez et al.
\cite{lop93}). Cliffe et al. (\cite{cli95}) studied the interaction
of a precessing jet with the interstellar medium. They find that
point-symmetric structures can be formed by this interaction, and
with the time, the individual bow-shocks can merge into a single
shock structure. The ulterior evolution of such structures will
lead to the formation of bipolar lobes with non-uniform brightness
distribution being point-symmetric respect to the central star.
On the other hand, Garc\'{\i}a-Segura \& L\'opez (\cite{gar00})
made models of bipolar PN with point-symmetric structures using a
steady misalignment of the magnetic collimation axis with respect
to the symmetry axis of the bipolar outflow. In their models, such
morphology is also produced by the action of jets.

A possible scenario for the formation of NGC\,6309 is that the knots
are the remnants of high-velocity bipolar collimated outflows that
were ejected in the proto-planetary nebula (PPN) stage. There is
evidence that precessing jets in young proto-PN carve bipolar or
multipolar cavities in the initial stage of PNe evolution
(e.g. Sahai \& Trauger \cite{sah98}, Sahai et al. \cite{sah05},
S\'anchez-Contreras et al. \cite{san06}). In such a scenario, the
precessing jet is ejected during the proto-PN phase and after that,
subsequently, it is stopped and cooled by its surroundings. The knots
follow the thermal expansion of the bipolar lobes. In this case, the
brightness of the knots comes mainly from the photoionisation process,
as the shocked-cooled emission has turned off or it is hidden by the
effect of a high-excitation photon source. This could explain the
lack of [\ion{N}{ii}] emission in the point-symmetric knots. 

\begin{figure}
  \resizebox{\hsize}{!}{\includegraphics{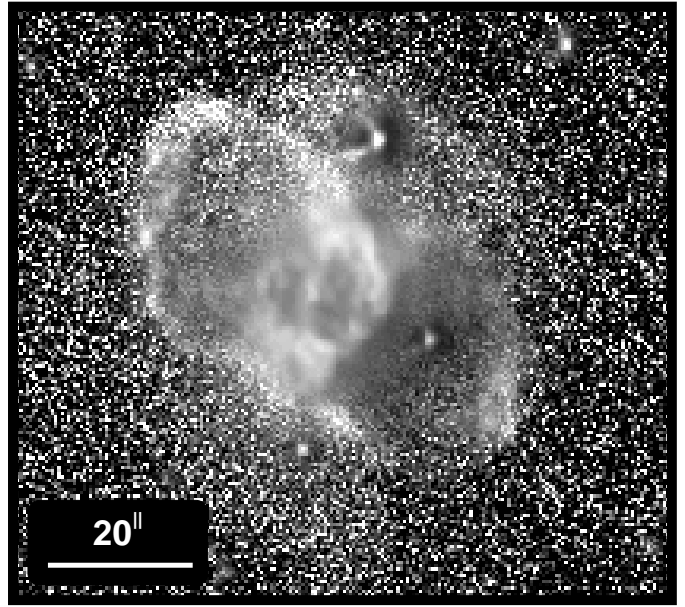}}
  \caption{[\ion{O}{iii}]/H$\alpha$ ratio image. White represents
   high values of the ratio.}
  \label{cociente}
\end{figure}

In addition, we constructed a [\ion{O}{iii}]/H$\alpha$ ratio image
from images in Fig.~\ref{cafos}. Such ratio image can be used as a
diagnostic tool, as has been shown by Medina et al. (\cite{med07}).
Figure~\ref{cociente} shows the [\ion{O}{iii}]/H$\alpha$ ratio image
in which an enhancement of the [\ion{O}{iii}] emission is seen in
the edge of the lobes, clearly related to the point-symmetric knots.
It is noticeable that the emission vanished at the end of the SW
lobe, possibly corresponding to a breaking of the shell due to the
action of the outflow. According to Medina et al. (\cite{med07}),
the [\ion{O}{iii}] enhancement would be compatible with structures
shaped by collimated outflows, as in the case of \object{IC\,4634}.
Finally, the circular halo probably corresponds to the remnant of
the envelope ejected as a slow wind from the central star when it
was at its AGB phase.

\section{Conclusions}

We have carried out an analysis of the planetary nebulae NGC\,6309, 
based on ground-based and space-based imaging, high- and
low-dispersion spectroscopy, as well as VLA-D radio continuum.
We summarize the main conclusions of this work as follows. 

NGC\,6309 can be described as a quadrupolar PN formed by a bright
central torus, two systems of bipolar lobes oriented at different
directions, and point-symmetric knots that trace the edges of the
lobes. The torus expands at 25\,km\,s$^{-1}$, whereas the polar 
expansion velocity of the lobes is 75\,km\,s$^{-1}$ for a first
bipolar system at PA 76\degr, and 86\,km\,s$^{-1}$ for another
bipolar system at PA 40\degr. Assuming a distance of 2\,kpc for
the nebula, the kinematic ages of the structures ranges from 3700
to 4000\,yr, suggesting that they have been formed in a short time
span. The knots at the edges of the lobes suggest that the lobes
have been formed by rapidly precessing bipolar jets that carved
cavities in the previous red giant envelope. In addition to these
structures, we detect a circular halo surrounding the torus and
bipolar lobes, which probably corresponds to the envelope ejected in
the AGB phase by the central star. There is also a conelike structure
embedded in one of the lobes and with its base on the torus.

We also study internal variations of the physical conditions and
chemical abundances in NGC\,6309. The low-dispersion spectra indicate
a high-excitation nebula, with low to medium variations of its
internal physical conditions 
 ($10,600\,{\rm K}     \lesssim T_{\rm e}$[\ion{O}{iii}] $\lesssim 10,900\,{\rm K}$; 
  $10,100\,{\rm K}     \lesssim T_{\rm e}$[\ion{N}{ii}]  $\lesssim 11,800\,{\rm K}$; 
  $1440\,{\rm cm}^{-3} \lesssim N_{\rm e}$[\ion{S}{ii}]  $\lesssim 4000\,{\rm cm}^{-3}$;
  $1700\,{\rm cm}^{-3} \lesssim N_{\rm e}$[\ion{Cl}{iii}]$\lesssim 2600\,{\rm cm}^{-3}$;
  $1000\,{\rm cm}^{-3} \lesssim N_{\rm e}$[\ion{Ar}{iv}] $\lesssim 1700\,{\rm cm}^{-3}$).
The radio continuum emission indicates a mean electron density of
$\simeq1900$\,cm$^{-3}$; emission measure of
$5.1\times10^5$\,pc\,cm$^{-6}$; and an ionised mass
$M$(\ion{H}{ii})$\simeq0.07\,$M$_\odot$. The logarithmic extinction
coefficient $c_{\rm H \beta}$ range from 0.70 to 0.97 (mean value
0.87), which probably is produced by differences in the internal
dust distribution. 

\begin{acknowledgements}
RV, SA, LO, MEC, and PFG were supported by grants 32214-E and 45848
(CONACYT), and by grants IN114199 and IN111903-3 (PAPIIT-DGAPA-UNAM).
LFM and SA were supported by grant AYA2005-01495 of the Spanish MEC
(cofunded by FEDER funds). JMT acknowledges partial financial support
from the Spanish grant AYA2005-08523-C03. We are grateful to the
staff of all the astronomical facilities used in this research,
namely: (a) Centro Astron\'omico Hispano-Alem\'an, (b) Very Large
Array of National Radio Astronomy Observatory, (c) Hubble Space
Telescope Data Archive, (d) Observatorio Astron\'omico Nacional,
operated by Universidad Nacional Aut\'onoma de M\'exico, and
(e) Observatorio de Sierra Nevada (IAA-CSIC). We thank the
anonymous referee for critically reading the manuscript and for
useful suggestions. We also thank fruitful discussions with
Dr. Mart\'{\i}n A. Guerrero (IAA-CSIC) and Prof. Mauricio Tapia
(IA-UNAM). This research has made use of the SIMBAD database,
operated at CDS, Strasbourg, France.

\end{acknowledgements}

\end{document}